\documentstyle[11pt,newpasp,twoside,epsf]{article}
\newcommand{\HI}{\protect\normalsize H\thinspace\protect\footnotesize I\protect\normalsize}

\def\edcomment#1{\iffalse\marginpar{\raggedright\sl#1\/}\else\relax\fi}
\reversemarginpar

\begin{document}
\title{The First Complete and Fully Sampled H\,{\small \bf I} Survey of the Tidal Arms of the Magellanic System}
 \author{Christian Br\"uns, J\"urgen Kerp}
\affil{Radioastronomisches Institut der Universit\"at Bonn, Auf dem H\"ugel 71,
D-53121 Bonn, Germany}
\author{Lister Staveley-Smith}
\affil{Australia Telescope National Facility, CSIRO, 
       PO Box 76, Epping NSW 1710, Australia}

\begin{abstract}
  The LMC, SMC, and the Milky Way are a spectacular set of interacting
  galaxies. The prominent tidal arms which protrude from the
  Magellanic Clouds extend for tens of kiloparsecs and cover a large
  part of the southern sky. These arms, and the Magellanic Clouds
  themselves, are therefore a highly suitable subject for the detailed
  investigation of an interacting galaxy system. We present the first
  results of the Parkes narrow-band \HI\ survey of the tidal arms of the
  Magellanic System. It is the first fully sampled, spatially complete
  survey of this system.  The data provide information on the column
  density distribution and the variation of the shape of the
  line profiles over the extent of the tidal features.  We present a
  cloud catalogue of the tidal features of the Magellanic System and
  discuss the distribution relative to the observational parameters
  longitude, latitude and radial velocity (v$_{\rm LSR}$ and v$_{\rm
    GSR}$).  We find a new stream-like feature that is most likely
  related to the Leading Arm. Detailed investigation reveals a number
  of clouds showing a cometary appearance in the position-velocity
  representation (so-called head-tail structures).  These head-tail
  structures are interpreted as clouds that are currently interacting
  with an ambient medium.
\end{abstract}

\section{Introduction}

Dieter (1965) surveyed the Galactic poles with the Harvard 60-ft
antenna and noted several detections with high radial velocities near
the southern Galactic pole which have no counterparts towards the
northern pole.  This was the first detection in \HI\ 21-cm line
emission of what we call today the ``Magellanic Stream'', but she did
not identify it as a coherent structure.  Further observations of
the northern part of the Magellanic Stream, called the
``south pole complex'' at this time, were done by Kuilenburg (1972)
and Wannier \& Wrixon (1972). Mathewson, Cleary, \& Murray (1974)
observed the southern part of the Magellanic Stream using the 18-m
Parkes telescope. They discovered the connection between the ``south pole
complex'' and the Magellanic Clouds and called it the ``Magellanic
Stream''.  Later on, several parts of the Magellanic Stream were
observed with higher resolution and sensitivity.

We now know that, next to the Milky Way/Sagittarius dwarf interaction,
the LMC, the SMC and the Milky Way form the nearest and certainly one of the
most spectacular examples of a galaxy interaction. Individual tidal
features that are traced by the \HI\ 21 cm line are: (1) the
Magellanic Bridge (MB), a high column density gas stream (N$_{\rm HI}
\geq 1 \times 10^{21}\,{\rm cm^{-2}}$ at 14' angular resolution)
connecting the stellar bodies of the LMC and SMC (see Fig. 1); (2) the
Magellanic Stream (MS), a coherent structure which starts from the
Magellanic Clouds and trails over 100$^{\circ}$ passing the southern
Galactic pole.  The column density in the bright parts of the MS is up
to N$_{\rm HI}\approx 5 \times 10^{20}\,{\rm cm^{-2}}$ at 14' angular
resolution; and (3) the Leading Arm (LA, Putman et al.~1998) that was
found to protrude from the Magellanic Bridge and the LMC along several
clumpy filaments. The existence of the LA was anticipated by 
interaction models but not previously detected as a coherent tidal
structure.

There are two major approaches to explain these features. The
first is a pure gravitational model (e.g.~Gardiner, Sawa, \& Fujimoto
1994), and the second is a ram-pressure model (e.g.~Moore \& Davis 1994).
The first model is able to explain the existence of a LA feature,
while the ram-pressure model fits the morphology of the MS much
better, but fails in explaining the LA. Recent models combine both
approaches by introducing drag forces to the gravitational model
(Gardiner 1999).

Recently, three large scale surveys in the southern sky have been made. The
first is the Argentinian survey (\HI\ Southern Sky Survey,
HISSS), presented by Bajaja et al.~in these proceedings.  It is
the counterpart to the Leiden/Dwingeloo survey and offers an angular
resolution of 30' with grid spacings of 30'. At a velocity resolution
of $\Delta v$ = 1 km ${\rm s}^{-1}$ the rms noise is $\sigma_{\rm rms}
\approx $ 0.07 K.  The second is the HIPASS survey, also presented in
these proceedings. It offers a velocity resolution of $\Delta v$ = 18
km ${\rm s}^{-1}$ and an rms noise of $\sigma_{\rm rms} \approx $ 0.01
K. The observing mode of the HIPASS is in-scan beam-switching. This mode
filters out all large-scale structure of the Milky Way or the
Magellanic System.  The northern part of the Magellanic Stream is not
(yet) complete (it is complete for $\delta_{2000} <$ +2\deg).  The
third survey is presented in this paper. Our survey was performed
using the narrow-band facility of the 64-m Parkes telescope. It is
fully sampled and covers the complete Magellanic System offering a
velocity resolution of $\Delta v$ = 0.8 km ${\rm s}^{-1}$ and an rms
noise of $\sigma_{\rm rms} \approx $ 0.12 K.

Our narrow-band \HI\ survey of the Magellanic System comprises
full sampling and high velocity resolution. Both are necessary for a
detailed investigation of the physical condition of the gas in the
Magellanic System and for improving the constraints which are
needed to improve the tidal and ram-pressure interaction models.

\section{Observations and Data Reduction}

\begin{figure}
\vspace*{-0.5cm}
\plotone{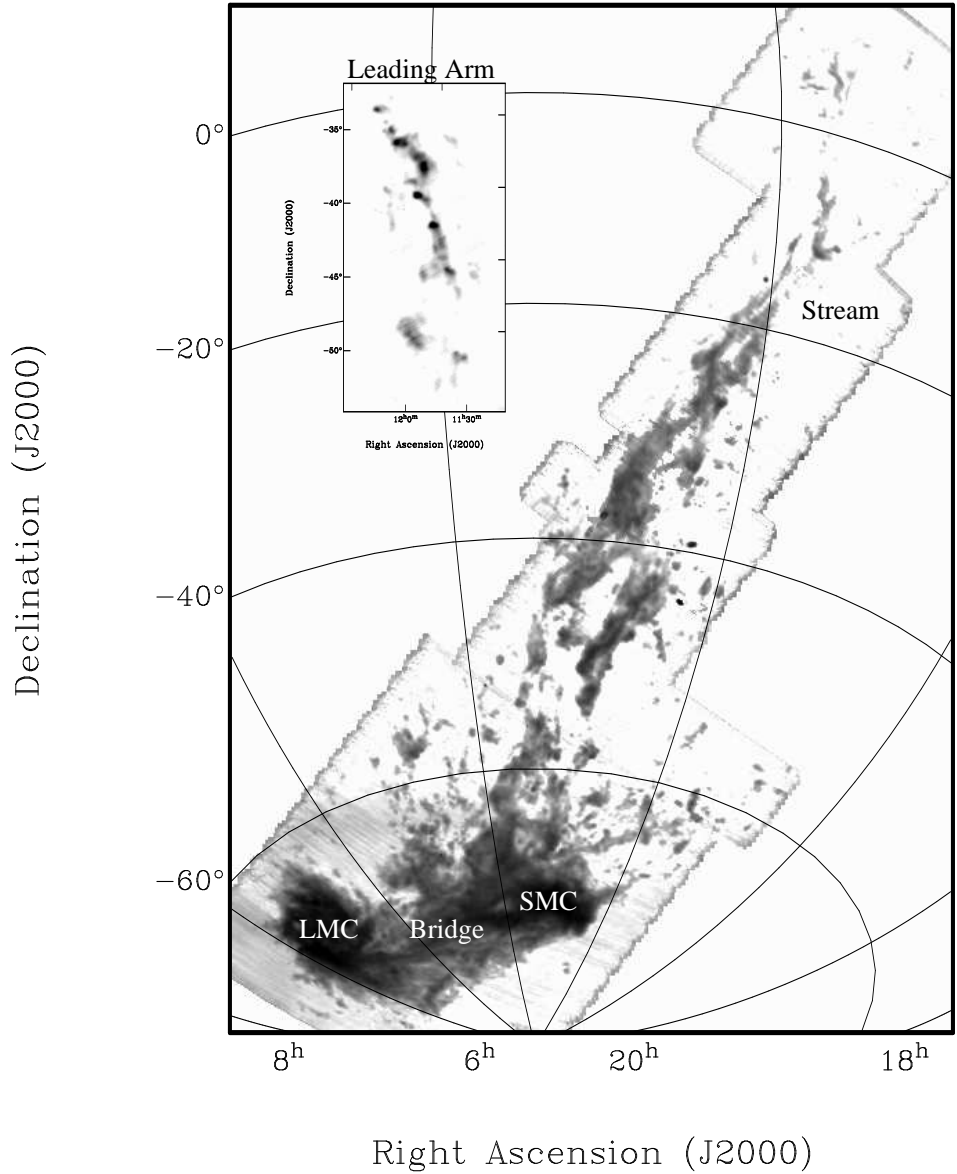}
\caption{\HI\ column density distribution of a part of the Parkes
narrow-band \HI\ Survey of the Magellanic System. The gray-scale has a
logarithmic intensity scale. The LMC and the SMC are on the lower left
part of the figure. The Magellanic Bridge is the high column density
connection between both galaxies. The Magellanic Stream starts near
the Bridge and continues to the upper right corner of the figure. The
inset shows the northern part of the Leading Arm feature. The
gray-scale for the inset is linear.}
\end{figure}

The observations were made within 10 days in Feb.~and Nov.~1999, using
the narrow-band multi-feed facility of the 64 m Parkes
radiotelescope. At 21 cm the telescope has a FWHP beamwidth of 14.1
arcmin.  The narrow-band system utilises the central 7 beams with two
orthogonal polarizations and a bandwidth of 8 MHz with a 2048 channel
autocorrelator for each beam and polarization. This configuration
allows in-band frequency-switching, i.e. the ``reference-spectrum''
also contains the spectral line. We used a frequency offset of 3.5
MHz. In-band frequency-switching doubles the integration time but
reduces the usable bandwidth to 4.5 MHz (950 km ${\rm s}^{-1}$).  The
\HI\ data were observed in on-the-fly mode with a telescope
scanning speed of 1\deg\ per minute. The integration time per spectrum
was set to 5 seconds for both the ``ON'' and the ``OFF''-spectra. The
feed-angle was set to 19.1\deg. The scans were aligned to Magellanic
coordinates. The offset between two adjacent scans was set to
0.56\deg.  The rms noise in the final data cubes is $\sigma_{\rm rms}
\approx $ 0.12 K at a velocity resolution of $\Delta v$ = 0.8 km ${\rm
  s}^{-1}$.

Because this survey is not a whole sky survey, we used the available parts of
the HIPASS survey to define the regions of interest with emission from the 
Magellanic Clouds and their tidal arms.

The data presented here are preliminary as there remain calibration, baseline
and interference problems to be removed. Nevertheless, the data are of high
quality and considerably improve on previously published observations.

\section{The Distribution of Magellanic Gas Clouds}

\begin{figure}
\vspace{-1.6cm}
\plotone{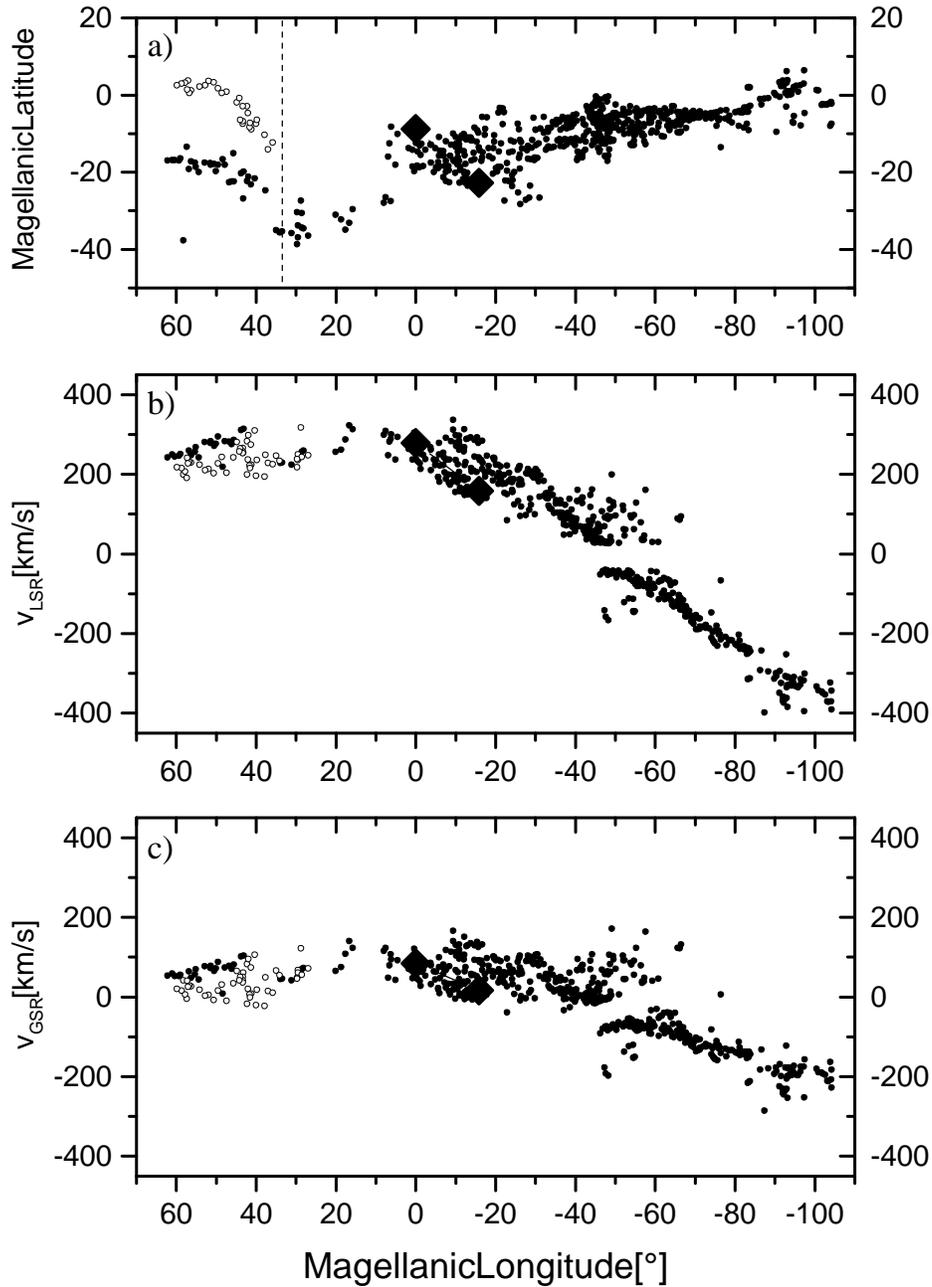}
\vspace {-0.8cm}
\caption{Each marker represents an entry in our cloud catalogue of the tidal 
arms of the Magellanic System. {\bf a}:
Magellanic latitude vs.~Magellanic longitude. {\bf b}: v$_{\rm LSR}$ 
vs.~Mag.~longitude. {\bf c}:  v$_{\rm GSR}$ vs.~Mag.~longitude. 
The Magellanic Clouds themself are marked by diamonds. The dashed line 
in the upper panel represents the location of the Galactic Plane. 
Filled circles are clouds that belong to the Magellanic Stream
or the Leading Arm as defined by Putman et al.~(1998). Open circles represent a
newly discovered feature that is most likely related to the Leading Arm 
feature.
}
\end{figure}

In this section we discuss the distribution of the \HI\ gas in the 
Magellanic System with respect to the observational parameters Magellanic 
longitude, latitude and radial velocity.

\subsection{General Distribution}
 
Figure 1 shows the column density distribution of the Magellanic Clouds, the
Magellanic Bridge (MB), the Magellanic Stream (MS), and a part of the Leading 
Arm (LA).
The velocity regime of the local gas (--25 km ${\rm s}^{-1} \le v_{\rm LSR} \le$ 
25 km ${\rm s}^{-1}$) was excluded to avoid confusion with
the Galactic gas.
In addition, in the northern regions of the Magellanic Stream 
($\delta_{2000} >$ --15$^\circ$) the velocity interval  
--80 km ${\rm s}^{-1} \le v_{\rm LSR} \le$ --25 km ${\rm s}^{-1}$ was excluded 
to avoid confusion with the intermediate velocity clouds (IVCs) in this area.
The IVCs are most likely associated with the Milky Way and not with the
Magellanic System.

The Magellanic Bridge is a complex structure having high column
densities (up to N(\HI)$~\approx$ 10$^{21}\,{\rm cm^{-2}}$ at
14' angular resolution).  The interface region between the Bridge and
the Stream is very complex. There is a lot of gas moving at different
velocities in different directions.  The Stream itself appears to be
much more confined. With the current resolution and sensitivity the
premise of the Stream consisting of six main clumps does not hold any
longer. Moreover, the main clumps contain several dense condensations
that appear partly as unresolved point sources. The Stream is divided
into two streams appearing like a twisted structure. This is possibly
due to the motion of the LMC and the SMC relative to each other,
producing a helix-like structure.  The column density decreases from
N(\HI) $\approx 5 \times 10^{20}\,{\rm cm^{-2}}$ at the
southern part of the stream to N(\HI) $\approx
1 \times 10^{19}\,{\rm cm^{-2}}$ at the northern tip.

The inset in Fig.1 shows the northern part of the Leading Arm. The LA
is a very narrow stream containing a number of dense condensations
that are not resolved with the Parkes telescope. In contrast to the
MS, the highest column densities in the LA are not found near the
Magellanic Clouds but in the most distant condensations. They show up
with the narrowest linewidth. In the MS, the most distant clouds show
up with the narrowest linewidth, too.  

\subsection{The Catalogue}

We have compiled a catalogue of clouds in the tidal arms of the
Magellanic System.  We exclude emission that is directly associated
with the LMC or the SMC. In addition, we exclude emission features in
the Magellanic Stream that have radial velocities similar to the local
gas of our own Galaxy (--25 km ${\rm s}^{-1} \le v_{\rm LSR} \le$ 25
km ${\rm s}^{-1}$).  In the region of the Leading Arm emission was
catalogued if it shows up with radial velocities v$_{\rm LSR} >$ 150
km ${\rm s}^{-1}$, because there are a number of HVCs at lower
velocities in this area that are most likely not associated with the
Magellanic System. 

A proper representation of the Magellanic System in common coordinates
is difficult, as both the southern poles of Equatorial and Galactic
coordinates fall into the region of interest.  Therefore, we use a
``Magellanic'' coordinate system\footnote{The Magellanic coordinates
  used in this paper are {\em not} the coordinates defined by Wannier \& 
  Wrixon (1972).}
as follows: the line of zero latitude is defined as the line from Galactic
longitude l = 90\deg\ across the southern Galactic pole to l =
270\deg. Zero longitude is defined by the position of the LMC.

\subsubsection{First results}

Figure 2 shows the catalogued distribution of the Magellanic gas clouds
with respect to the parameters Magellanic latitude (Fig. 2a), v$_{\rm
  LSR}$ (Fig. 2b), and v$_{\rm GSR}$ (Fig. 2c) as a function of
Magellanic longitude.  Each detected cloud (i.e. resolved clump) is
represented by a marker.  The Magellanic Clouds themself are marked by
diamonds. The dashed line represents the Galactic Plane. The filled circles
represent clouds that belong to the Magellanic Stream, the Magellanic
Bridge or the Leading Arm as defined by Putman et al.~(1998).  The
Magellanic Stream is connected to the Magellanic Bridge (see Figs. 1
and 2a).  In the interface region between Bridge and Stream, the gas
clouds have higher radial velocities than the gas located in the
Bridge and the SMC. The radial velocity changes dramatically from
v$_{\rm LSR} \approx $ +250 km ${\rm s}^{-1}$ near the Magellanic
Bridge to v$_{\rm LSR} \approx $ --400 km ${\rm s}^{-1}$ at the
northern tip. Figure 2c shows that this velocity gradient is much lower
in the Galactic-standard-of-rest frame, but is still significant ($\Delta
v \approx 3.5$ km s$^{-1}$ deg$^{-1}$).  While the Magellanic Stream
shows up with a large velocity gradient, the velocity in the Leading
Arm remains pretty constant in both v$_{\rm LSR}$ and v$_{\rm GSR}$.

The highest dispersion of the observed radial velocities is
observed near the southern Galactic pole. There are a number of gas
clouds having high velocities relative to the velocities of the
Magellanic Stream in the region where the Magellanic Stream passes the
line of sight to the Sculptor group (near Magellanic longitude
--50\deg).  There has been some debate as to whether the \HI\ clouds
with radial velocities comparable to the Sculptor galaxies are located
within the Sculptor group, or whether they are part of the Magellanic
Stream or ``normal'' Galactic High-Velocity Clouds (Mathewson et al.~1975).

The LA starts in the opposite direction of the MS. Near the
Galactic Plane it performs a 90$^\circ$ bend, possibly consistent with
deflection by a rotating, extended disk of the Galaxy. We plan to
examine this in more detail.  Open circles in Fig. 2 represent a newly
discovered feature that is most likely related to the LA. It is also a
stream-like structure but it is not as continuous as the ``old''
Leading Arm feature. The radial velocities of this new feature fit
very well to the LA defined by Putman et al.~(1998).

\begin{figure}
\plottwo{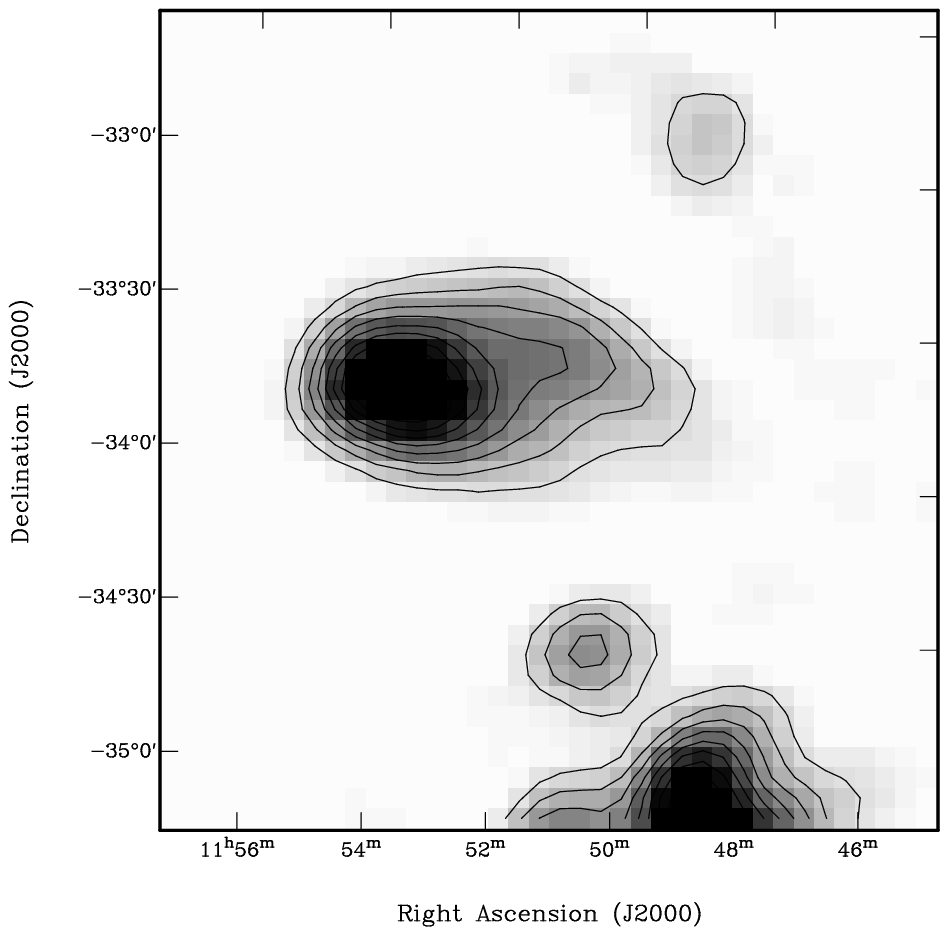}{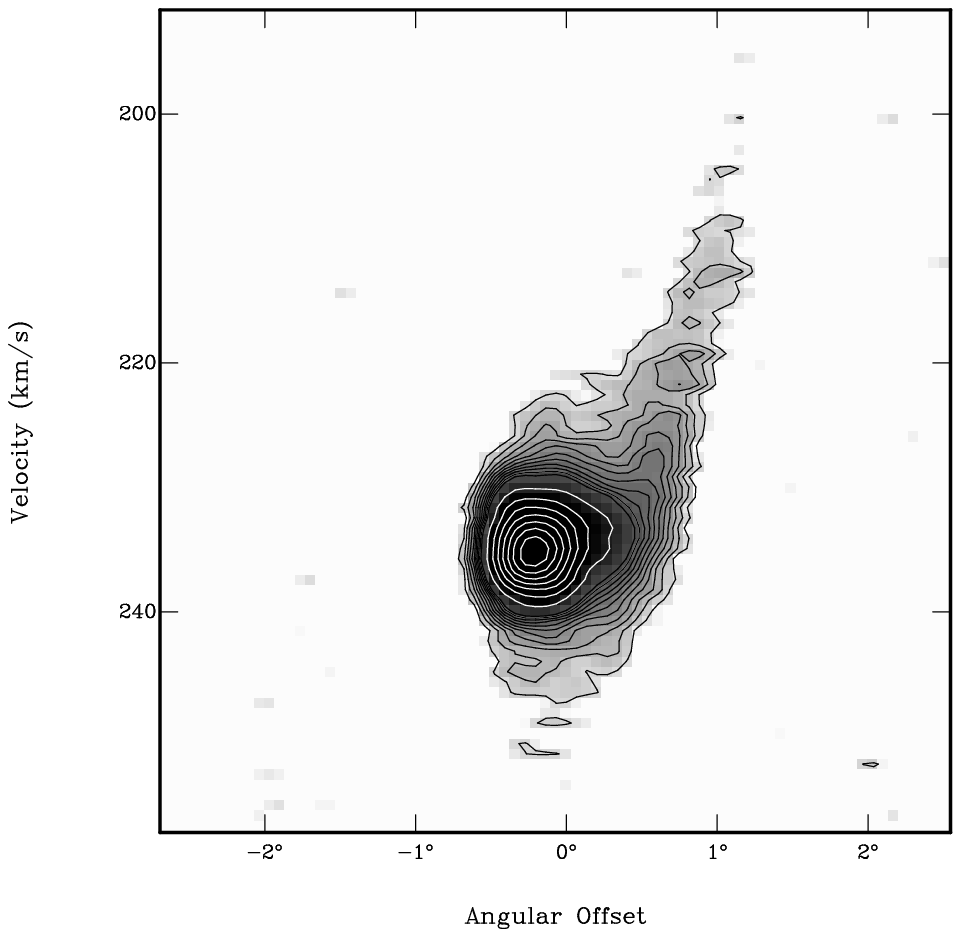}
\caption{An example of a head-tail cloud. {\bf right}: \HI\ column density 
distribution. There is a clear peak at the left side of the cloud and a
tail of lower column density to the right. {\bf left}: a slice through the 
data cube in the direction of the column density gradient (v$_{\rm LSR}$ vs. position).
Associated with the column density gradient is a significant velocity gradient.}
\end{figure}

\section{Head-Tail Structures}

The high velocity resolution ($\Delta$v = 0.8 km ${\rm s}^{-1}$) of
this \HI\ survey allows a detailed analysis of the shape of
individual line profiles.  Despite the fact that there are a lot of
baseline problems near the Galactic Plane, a detailed analysis of
individual clouds in the tidal features at high Galactic latitudes is
possible. A number of clouds have simultaneously a velocity and a
column density gradient. These clouds have a cometary appearance in
the position-velocity representation and are called henceforward
head-tail clouds (Br\"uns et al.~2000).  Figure 3 shows an example of a
head-tail cloud. The left figure shows the column density
distribution. There is a clear column density peak at the left side of
the cloud and a tail of lower column density to the right. The right
figure shows a slice through the data cube along the column density
gradient.  Associated with the column density gradient is a
significant velocity gradient.

A simple interpretation of the head-tail clouds is that they are
currently interacting with an ambient medium. In the context of this
model, the head (i.e. the high column density core of the cloud)
is the interacting cloud while the tails represent material that was
stripped off from the cloud's core.  Recent halo models (e.g.~Kalberla
\& Kerp 1998) predict very low densities in the outer halo (n
$\approx$ 10$^{-5}$ cm$^{-3}$ at a distance of 50 kpc). The existence
of head-tail clouds may be an indicator of an (at least locally)
increased halo density. Nevertheless, more detailed analysis of the
complete \HI\ survey and theoretical calculations are necessary
to reveal the nature of the head-tail clouds.

\section{Summary and Next Steps}

We have presented early results of the first spatially complete and
fully sampled \HI\ survey of the Magellanic System. We have also
compiled a catalogue of tidal gas clouds in the Magellanic System.
This catalogue may help to improve tidal models which attempt to reproduce the
morphology of the Magellanic System in detail.  We have discovered a
new stream-like structure that is most likely a part of the Leading Arm.
A future, detailed investigation of the small-scale structure within
the tidal arms may address whether there is any interaction with
the ambient medium.

\acknowledgments The authors thank P.M.W. Kalberla, U. Mebold, 
R. Haynes, M. Putman, M. Filipovic and E. Muller for their help during the
observations. C. Br\"uns thanks the
      \emph{Deut\-sche For\-schungs\-ge\-mein\-schaft, DFG\/} (project
      number ME 745/19) as well as the CSIRO for support.

\end{document}